\documentclass[fleqn,10pt]{wlscirep}
\usepackage[utf8]{inputenc}
\usepackage[T1]{fontenc}
\title{Spontaneous Pattern Formation in Intertype Superconducting Films}

\author[1]{W. Y. C\'{o}rdoba-Camacho}
\author[2]{R. M. da Silva}
\author[2,*]{A. A. Shanenko}
\author[3]{A. Vagov}
\author[1,4]{A. S. Vasenko}
\author[1]{B. G. Lvov}
\author[2]{J. Albino Aguiar}
\affil[1]{National Research University Higher School of Economics, Moscow, 101000, Russia}
\affil[2]{Universidade Federal de Pernambuco, Departamento de F\'isica, Recife, 50740-560, PE, Brazil}
\affil[3]{Institut f\"{u}r Theoretische Physik III, Bayreuth Universit\"{a}t, Bayreuth, 95440, Germany}
\affil[4]{Donostia International Physics Center (DIPC), Paseo Manuel de Lardizabal 4, San
Sebasti\'{a}n/Donostia, 20018 Basque Country, Spain}
\affil[*]{arkadyshanenko@df.ufpe.br}


\begin{abstract}
Thin superconducting films are usually regarded as type II superconductors even when they are made of a type I material. The reason is a strong contribution of the stray magnetic field that stabilizes vortices. While very thin films indeed reach this limit, there is a large interval of film thicknesses where the magnetic properties cannot be classified as either of the two conventional superconductivity types. Recent calculations revealed that in this interval the
system exhibits spontaneous formation of complex condensate-field patterns that are very sensitive to system parameters, in particular, the temperature and the applied magnetic field. The corresponding superconducting magnetic properties can be attributed to a special regime of the intertype superconductivity whose physical origin lies in the removal of an infinite degeneracy of the self-dual superconducting state at the critical Bogomolnyi point. Here we demonstrate that qualitative characteristics of the intertype superstructures in thin superconducting films are independent of the choice of the in-plane boundary conditions for the order parameter and the magnetic field.
\end{abstract}
\begin{document}

\flushbottom
\maketitle
\thispagestyle{empty}

\section*{Introduction}

It has been demonstrated, both experimentally and theoretically, that the traditional dichotomic type I - type II classification of the superconductor magnetic response is incomplete even in the simplest case of a clean BCS bulk superconductor~\cite{Krageloh,Essmann,Aston,Kumpf,Jacobs,Hubert,Auer,Klein,Weber,Brandt,Luk,Miran,Laver, Muhl,Brandt1,Pau,Reim,Ge1,Ge2} and has to be amended by an additional regime that can be referred to as the intertype (IT) superconductivity~\cite{Vagov,Wolf1,Wolf2}. This regime is characterized by the non-standard magnetic properties due to unconventional condensate-field spatial configurations in the mixed state. It is a generic phenomenon~\cite{Vagov} closely connected to the self-duality and the related degeneracy of the condensate-field state at the critical Bogomolnyi point (B point) ~\cite{Bogomolnyi1,Bogomolnyi2}. The IT superconductivity is found in the domain in the vicinity of the B point where the degeneracy is removed. There are many mechanisms for the removal, for example, induced by the non-local interactions in the condensate below the critical temperature $T_c$ or by the geometry related factors in finite and low-dimensional superconducting samples~\cite{Cordoba1,Cordoba2}. Furthermore, the IT superconductivity can be strongly enhanced by the presence of many conduction bands~\cite{Vagov} and also near the BCS-BEC crossover~\cite{Wolf1}.

It was recently suggested that the IT regime can be also achieved in thin superconducting films, made of a type-I material, where superconductivity is modified due to the strong influence of stray magnetic fields outside the sample~\cite{Cordoba1}. It is long known that the stray fields give an additional repulsive contribution to the interactions between vortices, which results in that an ultra-thin type-I superconducting film demonstrates a type-II behavior~\cite{Tinkham,Pearl,Maki}. This gave rise to a tacit assumption that all sufficiently thin superconductors exhibit a type-II behavior. However, more involved studies of type-I superconducting films have also demonstrated the existence of giant (multi-quantum) vortices~\cite{Lasher, Dolan, Hasegawa, Sweeney, Palonen, Gladilin} not inherent in type-II superconductors. A recent systematic analysis~\cite{Cordoba1} has revealed a finite interval of thicknesses at which type-I films produce unconventional mixed-state configurations that cannot be attributed to either of the standard superconductivity types. Instead of routine lamellas, typical for type I, or Abrikosov lattices of single-quantum (Abrikosov) vortices, expected in type II, these films exhibit spontaneous formation of diverse and rather complex patterns. At least three basic patterns can be distinguished: 1. superstructures made of the condensate bubbles surrounded by vortices; 2. the condensate stripes separated by vortex chains/labyrinths; and 3. mixtures of giant vortices and vortex clusters. Examples of systems with the spontaneous pattern formation are well-known in the literature and include magnetic films~\cite{Seul1, Seul2}, liquid crystals~\cite{Mac}, multilayer soft tissues~\cite{Seul2, Stoop}, lipid monolayers~\cite{Kell}, granular media~\cite{Aran} etc. Recently stripes of vortices have attracted interest in the context of unusual mixed-state configurations in ${\rm MgB}_2$~\cite{Moshch, Bend}. The findings of Ref.~\cite{Cordoba1} has demonstrated new and rather complex examples of self-organized patterns in superconductivity (e.g., with the coexistence of the condensate textures and vortex superstructures) and can be interpreted as the IT regime in thin superconductors that is similar to the IT superconductivity in bulk materials~\cite{Vagov, Wolf1, Wolf2}.

The focus of the earlier systematic study~\cite{Cordoba1} was thin films with very large in-plane dimensions (far beyond the mesoscopic regime). However, in practice the analysis employed the periodic boundary conditions for the condensate density and magnetic field. This raises a question whether the exotic configurations can be artefacts of such periodic boundary conditions. Indeed, the fixed boundaries for the in-plane supercurrent in real flat samples can have a strong influence on the superconducting state due to the boundary-condensate interactions (the Bean-Livingston barrier~\cite{BL}) and significant modifications in the stray magnetic field. The boundary-induced modifications of the stray fields decay relatively slow (non-exponentially) along the film and thus cannot be neglected even at sufficiently large distances. 

In addition, the periodic boundary conditions modify the mixed state via that in this case the total magnetic flux through the film is the product of the external field and the film surface area (due to the flux conservation). Thus, an arbitrary small applied magnetic field penetrates the sample and the Meissner state cannot be achieved. In turn, the mixed-state properties near the point of the transformation from the Meissner to the mixed state cannot be investigated. In this respect the case of a finite flat sample is qualitatively different. Finally, the influence of the boundary conditions can be amplified in the vicinity of the B point, where all condensate-field configurations in the mixed state are close to the degeneracy and even small boundary-induced perturbations can result in considerable changes of the condensate state. 

This work investigates details of the type I - type II crossover in type-I films of finite dimensions. The analysis takes into account stray magnetic fields outside the sample - the source of the crossover. The main goal is to complement the previous study of infinite films, performed for the periodic geometry~\cite{Cordoba1}, in order to reveal how the choice of the in-plane boundary conditions affects the intertype condensate-field patterns. 

\begin{figure}[]
 \includegraphics[width=0.7\linewidth]{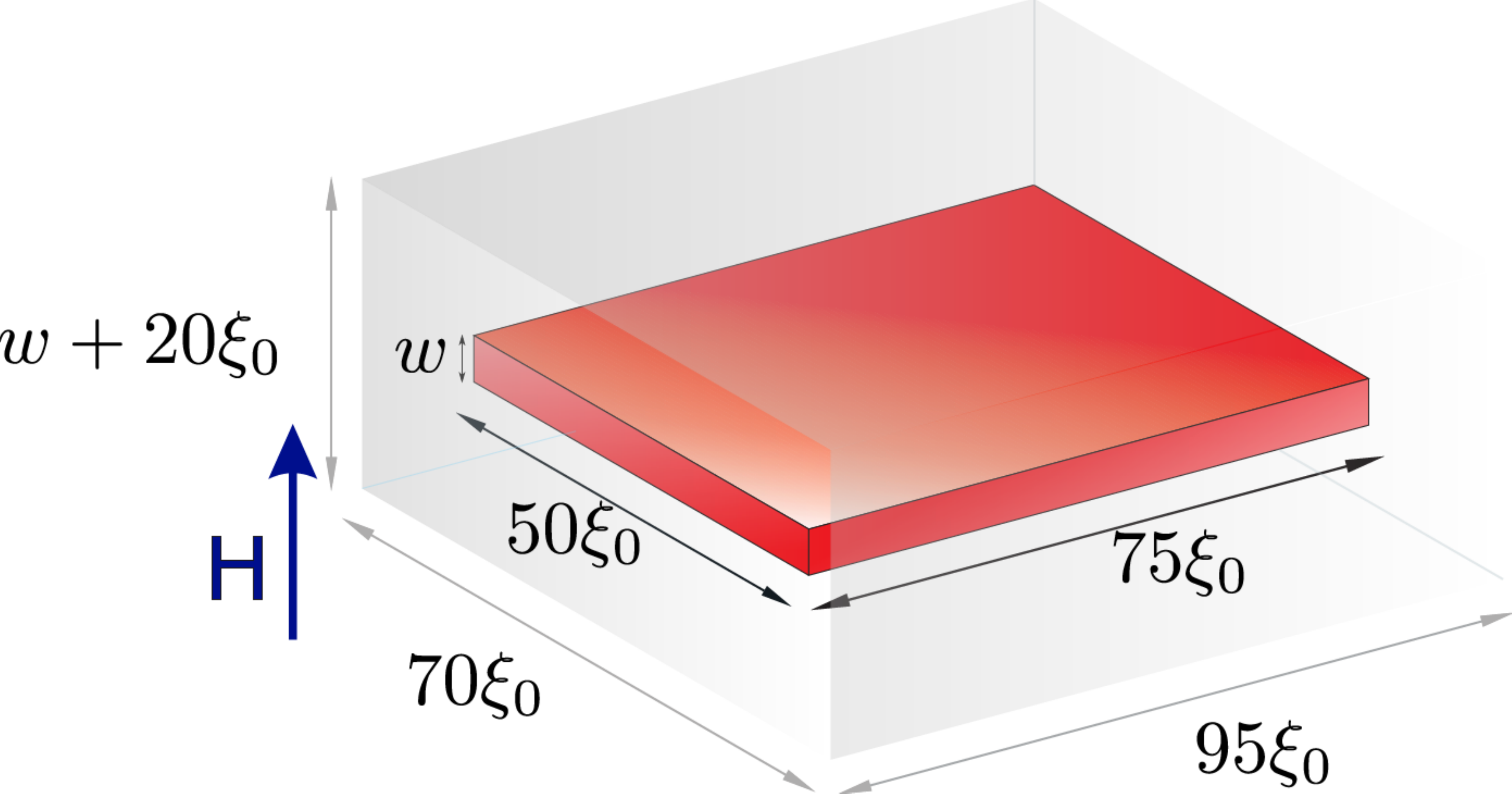}\hfil
\caption{A sketch of the superconducting slab (marked red) placed in the external perpendicular uniform magnetic field $\bf H$. In the grey area the field differs substantially from the uniform external magnetic field.} 
\label{fig1}
\end{figure}

To avoid any misunderstanding, we stress that the present consideration does not concern mesoscopic superconductors, extensively discussed previously~\cite{Schwei,Yamp,Cabral,Geurts}, where superconducting vortices are directly restricted by the sample boundaries. In our work the in-plane dimensions of a flat sample exceed the vortex core diameter considerably, as seen from the sketch in Fig.~\ref{fig1}. Thus, the condensate state is affected by its interaction with the stray fields that is controlled by the film thickness.

\section*{Results}

The analysis is done using the approach based on the GL theory as described in Methods section~\ref{sec_methods}.  Results of the study are illustrated by colour density plots of the squared absolute value of the order parameter $|\Psi|^2$ (the condensate density) in Figs.~\ref{fig2} and \ref{fig3} that show the condensate density at the central plane inside the superconducting sample. Figure~\ref{fig2} illustrates the temperature dependence of the condensate configuration calculated for several selected thicknesses of the film at a chosen value of the external magnetic field. In addition, Fig.~\ref{fig3} demonstrates how the condensate spatial distribution changes with the applied magnetic field at a chosen temperature.

\subsection{Temperature dependence}

\begin{figure*}[]
 \includegraphics[width=1.0\linewidth]{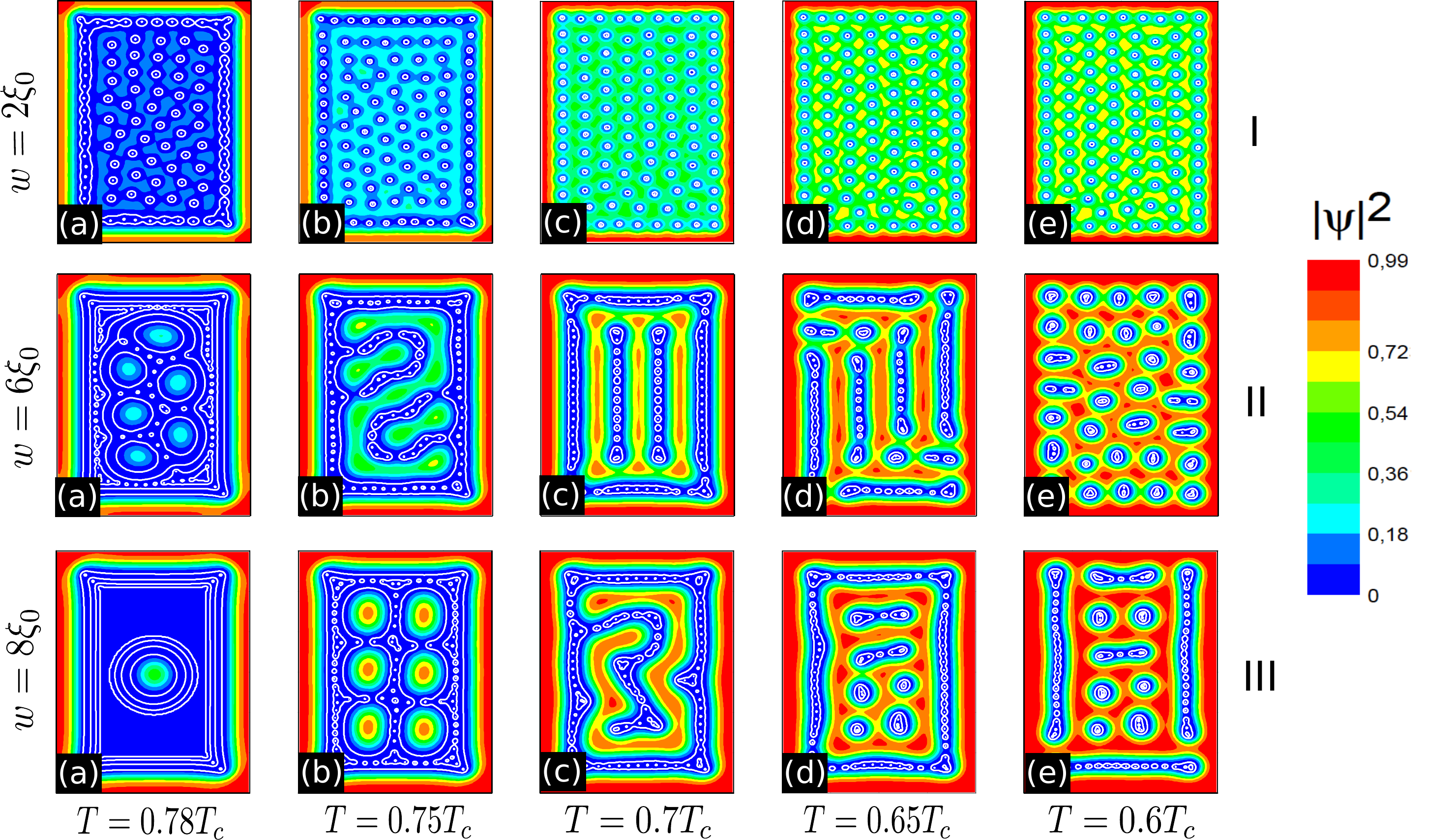}\hfil
\caption{The local density of Cooper pairs $|\Psi |^2$~(relative to its maximal value) calculated for different film thicknesses $w= 2\xi_0$ (upper row of panels - I), $w= 6\xi_0$ (middle row of panels - II), and $w=8\xi_0$ (lower row of panels - III). Panels correspond to different temperatures shown below. The color scheme is given on the right side and the scaled density varies from $|\Psi|^2 = 1$~(red) to $|\Psi|^2 = 0$~(blue).  The calculations are done for the external field $H=0.2 H_c(0)$. } \label{fig2}
\end{figure*}

Figure~\ref{fig2} shows the condensate spatial profile for the film thicknesses $w = 2\xi_0$, $6\xi_0$, and $8\xi_0$ ($\xi_0$ is the zero-temperature coherence length in bulk), placed in the perpendicular magnetic field $H=0.2 H_c(0)$ [$H_c(0)$ is the bulk thermodynamic critical magnetic field at $T=0$], at the temperature in the range $T=0.6$-$0.78T_c$. Notice, that critical temperature  $T_c$ corresponds to zero magnetic field. The material Ginzburg-Landau (GL) parameter is taken as $\kappa = 0.55$~($\kappa = \lambda/\xi$, with $\lambda$ the magnetic penetration length and $\xi$ the GL coherence length). 

The condensate configurations of the thinnest film, with $w=2\xi_0$, are typical for type II superconductors: magnetic field penetrates the sample in the form of well separate Abrikosov vortices, arranged in a distorted triangular Abrikosov lattice. The distortion is mainly a consequence of the incommensurability between the lattice and sample geometries. It increases at higher temperatures because of the interplay between a finite accuracy of numerical calculations and the near-degeneracy of the vortex configurations caused by the proximity to the B point. Another reason for this increase is slowing down of the numerical-procedure convergence when approaching the normal-to-superconducting transition. One sees that the superconductivity is enhanced in the vicinity of the boundaries, which is typical for finite type-II superconductors in a magnetic field. One concludes that for $w=2\xi_0$ the stray fields yield a major contribution to the vortex-vortex interaction - initially attractive vortices (in a bulk type-I material) become repulsive and form an Abrikosov lattice, in agreement with earlier works on superconducting films~\cite{Tinkham,Pearl,Maki}.

For thicker samples, $w = 6\xi_0$ and $8\xi_0$, the condensate profile is dramatically different from what is expected for both conventional superconductivity types. Moreover, it is very sensitive to the temperature, undergoing qualitative transformations when changing $T$. At $T=0.78T_c$ the field occupies almost all area of the sample with the exception of several superconductive islands of almost circular shape [Figs.~\ref{fig2} II(a) and III(a)]. The islands are arranged in superlattice structures, so that the situation is somewhat opposite to the Abrikosov lattice in type-II superconductors. At $T=0.78T_c$ the superlattice is visible for the thinner sample with $w=6\xi_0$, [Fig.~\ref{fig2} II(a)], while for the thicker one with $w= 8\xi_0$, it develops only at the lower temperature $T=0.75T_c$ [Fig.~\ref{fig2} III(b)]. One can see also vortices (white empty circles) in the domain of an almost suppressed condensate (blue background).

When the temperature decreases, the condensate profile changes qualitatively for both $w = 6 \xi_0$ and $8 \xi_0$. One sees that superconducting and nearly-normal regions are extended into quasi-one dimensional structures. They are arranged in large-scale superstructures that tend to reflect the sample symmetry [Figs.~\ref{fig2} II(b), II(c) and III(c)]. Domains with a strongly suppressed condensate are densely populated with vortices that form vortex chains. 
The rearrangement of superconductive islands into stripes separated by vortex chains is most pronounced at $T=0.7T_c$. 

At lower temperatures $T=0.65T_c$ and $0.6T_c$, longer vortex chains break into shorter ones, eventually forming separated vortices and vortex clusters. The breaking starts in the middle of the sample whereas close to its boundaries larger chains still remain [Figs.~\ref{fig2} II(d), III(d) and III(e)]. At $T=0.6T_c$ all chains in the thinner film with $w=6\xi_0$ are split into separate vortices and vortex clusters [Fig.~\ref{fig2} II(e)] while for $w=8\xi_0$ vortex chains still survive near the boundaries [Fig.~\ref{fig2} III(e)]. It should be noted that at this stage most of vortices carry multiple magnetic flux quanta. 

One anticipates further fragmentation of the vortex matter at lower temperatures, so that multiquantum vortices and vortex clusters finally become Abrikosov vortices arranged in an Abrikosov lattice. However, since the use of the GL theory is questionable at very low temperatures, the corresponding results are not shown here. 

\subsection{Field dependence}

\begin{figure}[]
 \includegraphics[width=0.7\linewidth]{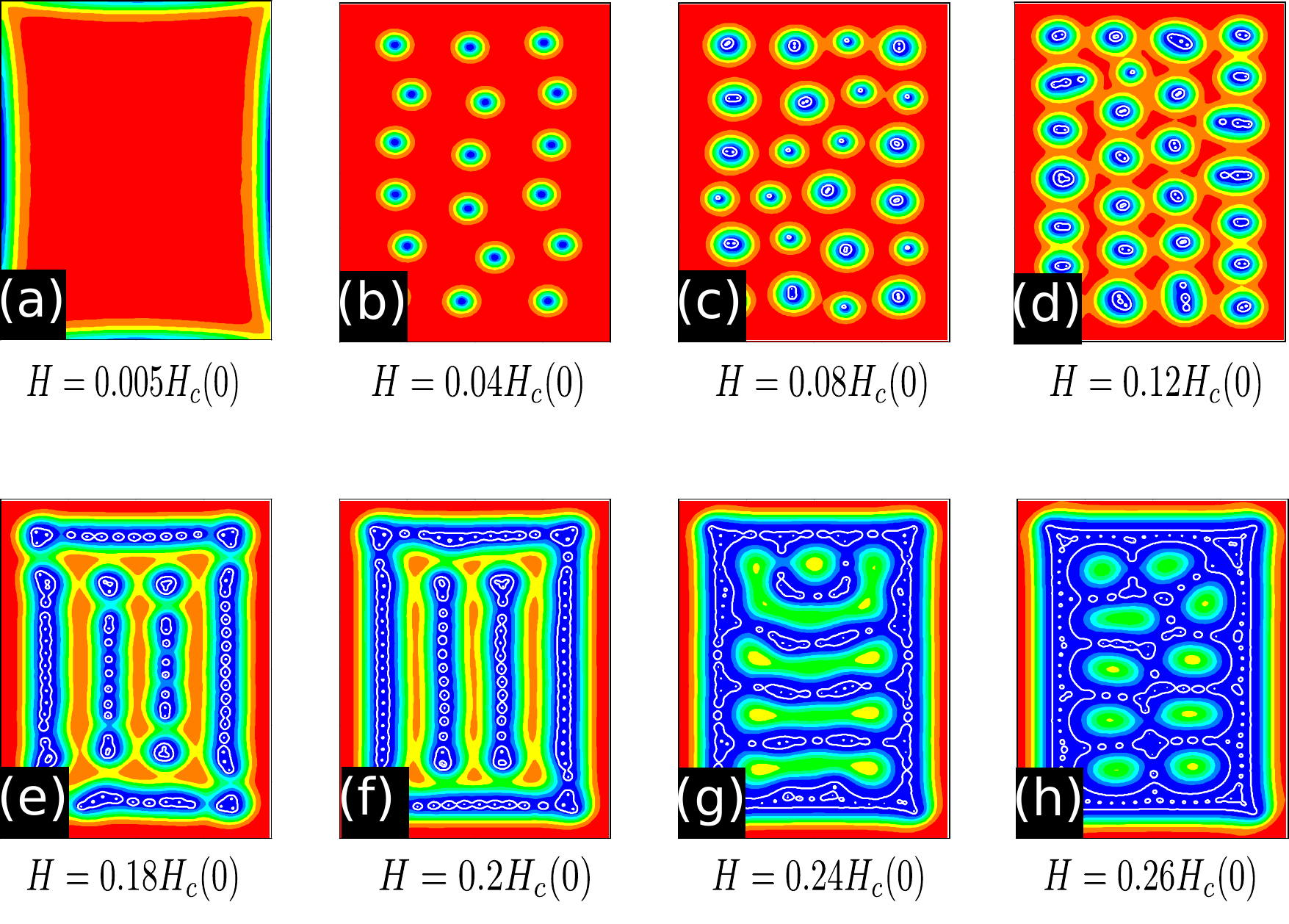}\hfil
\caption{Panels (a)-(h) represent the local condensate density $|\Psi |^2$ calculated for different values of the external magnetic field (shown below). Calculations are performed for $T=0.7 T_c$ and $w=6\xi_0$.  } 
\label{fig3}
\end{figure}

A further insight into changes of the condensate configurations is obtained by tracing their dependence on the magnetic field. Figure ~\ref{fig3} shows spatial profiles of the condensate density calculated for $w=6\xi_0$, $T=0.7 T_c$ and for the magnetic field in the range $H = 0.005$-$0.26 H_c(0)$.

At the lowest field, $H=0.005H_c(0)$, the system demonstrates the usual Meissner state [Fig.~\ref{fig3} (a)], where the field penetrates only vicinities of the boundaries. When $H$ increases, the field starts to enter the inner part of the sample in the form of Abrikosov vortices [Fig.~\ref{fig3} (b)] and the Meissner state is replaced by the non-uniform mixed state. The condensate is no longer suppressed in the vicinity of the boundaries.

A further increase of the field gives rise to the appearance of multiquantum vortices [Fig.~\ref{fig3} (c)] and vortex clusters [Fig. \ref{fig3} (d)]. Then vortex clusters start forming chains as shown in Fig.~\ref{fig3} (d) [cf. Fig.~\ref{fig2} II(e)]. The vortex chains grow in length eventually forming superstructures that reflect the system geometry [Fig.~\ref{fig3} (e), see also Figs.~\ref{fig2} II(d) and II(c)]. At still larger fields the regions of strongly suppressed condensate grow and eventually form domains of nearly normal state that surround isolated superconducting islands [Figs.~\ref{fig3} (g) and (h)]. Notice, however, that these nearly normal domains retains a vortex structure, similar to Figs.~\ref{fig2} II(a) and III(b).

\begin{figure}[]
 \includegraphics[width=0.7\linewidth]{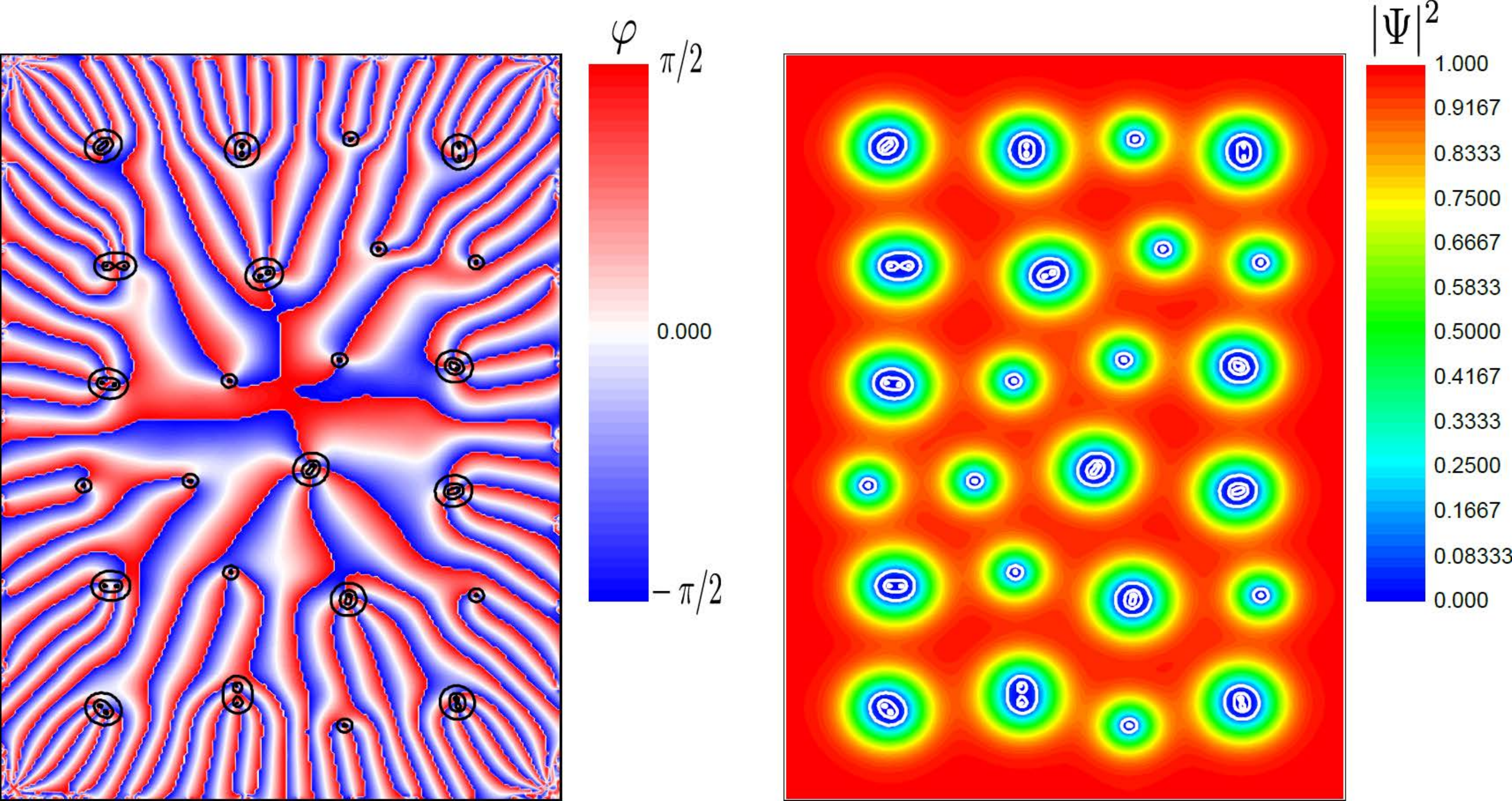}\hfil
\caption{Phase portrait (left panel) corresponding to the spatial profile of the condensate density with multiquantum vortices (right panel), as calculated for the sample with $w= 6\xi_0$ at $T=0.7T_c$ and $H=0.08H_c(0)$. The colour scheme for the phase portrait is dichromic - blue for $-\pi/2<\phi <0$ and red for $\pi/2> \phi >0$. } \label{fig4}
\end{figure}

The appearance of the multiquantum vortices can be seen from the phase portrait of the condensate profile calculated for the film with $w = 6\xi_0$, $T=0.7T_c$, and $H = 0.08 H_c(0)$, see Fig.~\ref{fig4} [cf. Fig.~\ref{fig3} (c)]. Here the phase distribution exhibits many vortex centers, where the phase changes its sign four times around the vortex center. This means that the corresponding phase accumulation is $4\pi$, which corresponds to two magnetic flux quanta. One can also see vortex clusters, made of two single-quantum vortices, in Fig.~\ref{fig4}.

\begin{figure}[]
 \includegraphics[width=0.7\linewidth]{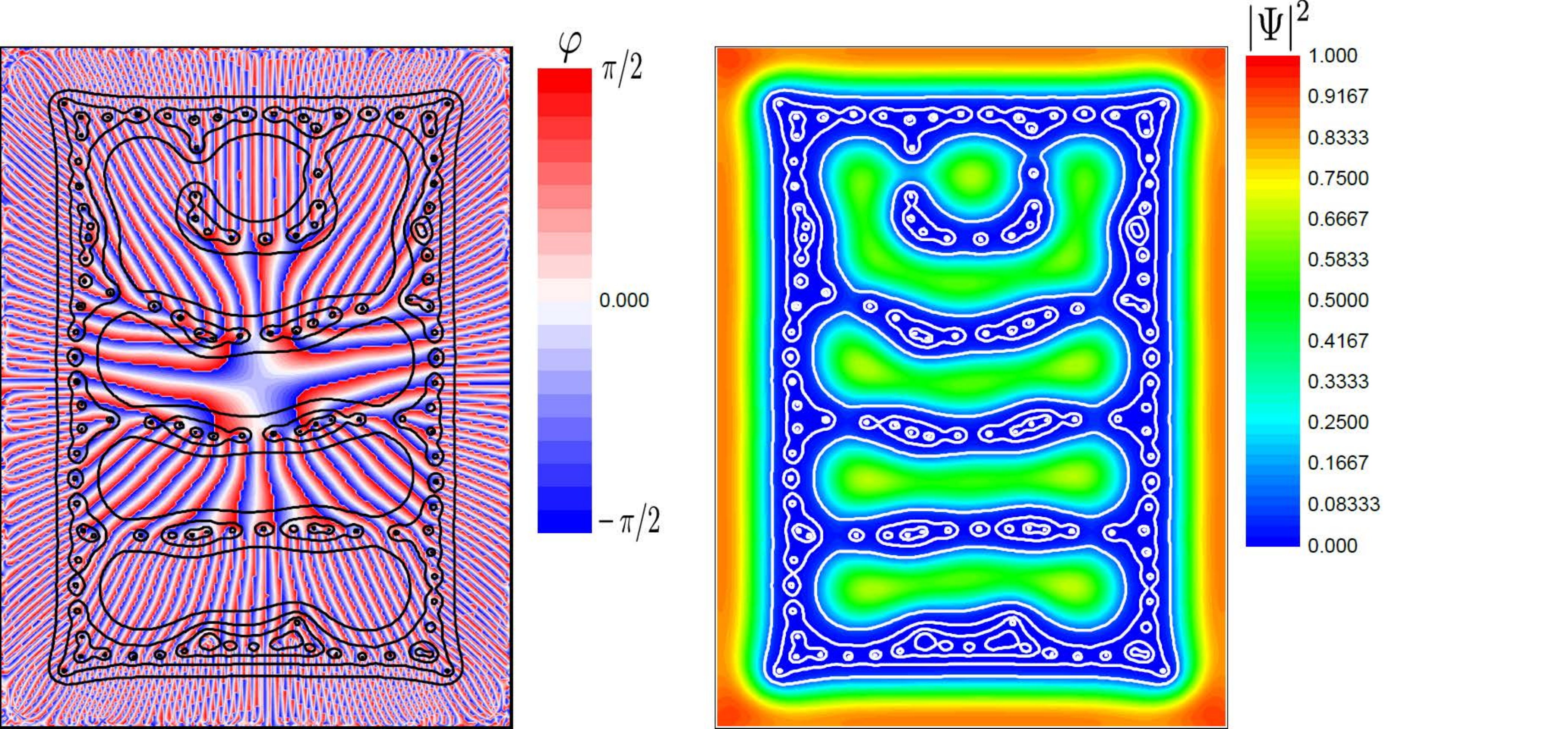}\hfil
\caption{The same as in Fig.~\ref{fig4} but for the applied field $H=0.24H_c(0)$, at which the stripe/bubble condensate pattern appears.} \label{fig5}
\end{figure}

In addition, Fig.~\ref{fig5} [cf. Fig.~\ref{fig3} (g)] demonstrates the spatial distribution of the order-parameter phase for the stripe pattern that appears in the film with $w = 6\xi_0$ at $T=0.7T_c$ and $H = 0.24 H_c(0)$. One can clearly see chains of single-quantum vortices separating the condensate stripes-islands.  

The dependence on the magnetic field can be thus summarized as follows: at sufficiently low fields one observes standard Abrikosov vortices, which at higher fields are sequentially transformed into multiquantum vortices, vortex clusters and vortex chains (labyrinths). These transformations of the vortex distribution are accompanied by the formation of the condensate superstructure made of stripes and islands (bubbles). Comparing Figs.~\ref{fig2} II, III and \ref{fig3} one sees that changes in the mixed-state configuration induced by increasing both the temperature and magnetic field are qualitatively similar.
 
\section*{Summary and Discussion}

This work studies superconducting magnetic properties of thin films with finite in-plane dimensions. The focus is the crossover from type-I to type-II superconductivity in films made of a type-I material. It is demonstrated that within a certain range of the film thickness [$2\xi_0 \lesssim w \lesssim 10 \xi_0$ for $\kappa = 0.55$] spatial configurations of the mixed state differ considerably from what is expected in both types I and II. In this thickness range we observe the spontaneous formation of complex condensate-field patterns that are found to be very sensitive to the system parameters. A general sequence of qualitative transformations of the mixed state is identified when one increases the external field and the temperature. 

Near the normal-to-superconducting transition the film reveals superconducting islands that are arranged in superstructures and surrounded by large domains of a suppressed condensate with vortices inside. When the temperature decreases such domains transform into closed-packed vortex chains while the islands are extended and become  superconducting stripes. Then, starting from the inner part of the sample, the chains break into vortex clusters and separated vortices,  most of which carry several magnetic flux quanta. Finally, when approaching the Meissner state, single-quantum vortices appear and tend to arrange themselves in an Abrikosov lattice. 

These patterns and their transformations are qualitatively similar to those obtained earlier for infinite type-I films with the in-plane periodic boundary conditions~\cite{Cordoba1}. Consequently, the qualitative picture of the mixed state in the IT superconducting films is not much affected by the boundary conditions and, therefore, the earlier results are not artefacts of the periodic geometry. However, the interaction with the in-plane boundaries of a finite sample is noticeable so that the superstructures of the condensate and vortices tend to follow the sample geometry. 

The observed transformations of the condensate-field configurations are apparently more complex than what is suggested by the simplified type II/1 concept of the IT superconductivity, see Ref.~\cite{Auer}. While the type II/1 concept is based on the long-range attraction between vortices, in our case the stray-field contribution results in a long-range repulsion of vortices. At short distances the vortex-vortex forces are attractive due to the type-I material of the sample. It is known that stripe patterns appear usually due to the competition between short-range attraction and long-range repulsion [see, e.g., Refs.~\cite{Seul2,Stoy}], which agrees with the appearances of the vortex chains in this work. In addition, recent calculations for the IT domain in bulk materials have demonstrated that the many-body (many-vortex) interactions~\cite{Wolf2} are responsible for the formation of vortex clusters and multiquantum vortices. Similar clusters and giant vortices are seen in Figs.~\ref{fig1} and \ref{fig2}, which makes it possible to expect that many-vortex interactions can also play the role here. 

Finally, we note that qualitatively similar mixed-state configurations have been recently observed in the calculations for superconducting nanowires~\cite{Cordoba2}. A direct comparison of the IT configurations in films and wires on one side and of bulk superconductors on the other is not possible at present because a detailed investigation of the mixed state in bulk IT superconductors is not yet available. However, it is clear that the spontaneous pattern formation in the IT superconductivity regime is directly connected with the removal of the degeneracy of the self-dual superconducting condensate state that takes place in the vicinity of the critical B point.  This represents a completely new mechanism of the self-organized  modulated phase structures.

\section*{Methods} 
\label{sec_methods}

The analysis is done using the GL theory. It is known that the IT superconductivity in bulk samples cannot be described within the GL approach~\cite{Vagov}. Within the GL theory the IT regime reduces to the single temperature independent point $\kappa=\kappa_0= 1/\sqrt{2}$. At this point the superconducting state is characterized by the Bogomolnyi self-duality~\cite{Bogomolnyi1,Bogomolnyi2}, which gives rise to the infinite degeneracy of all condensate-field configurations (including exotic superstructures and patterns)~\cite{Vagov,Wolf1,Wolf2,Weinberg}. 

Corrections to the GL theory, which take into account non-local interactions in the condensate and contribute at $T < T_c$, remove the degeneracy so that the critical point $\kappa=\kappa_0$ unfolds into a finite temperature-dependent interval of $\kappa$'s, forming the IT domain in the $\kappa$-$T$ plane, where the mixed state exhibits unconventional field-condensate patterns. When $T \to T_c$, the IT domain is reduced to the critical B point $(\kappa_0,T_c)$ at which the superconducting state remains self-dual and infinitely degenerate.

 The B-point degeneracy can also be removed by other physical mechanisms, including the interaction of the condensate with the sample boundaries and stray fields. These geometry-related mechanisms can be captured within the GL approach and the latter can be consequently employed to investigate the IT regime in thin superconducting films and wires.

In the calculations we solve the GL equations, which are written in the dimensionless units~\cite{Cordoba1, Cordoba2} as
\begin{align}
\big( - i \mathbf{\nabla} - {\bf A} \big)^2  \Psi  - (1-T) \big( 1 - | \Psi | ^2 \big)    \Psi  = 0,\quad 
\kappa^2 \, {\rm rot} \,  {\bf B}  =  (1-T) \, {\bf Re} \big[\Psi^\ast  \big( - i \mathbf{\nabla} - {\bf A} \big)^2 \Psi   \big] ,
\label{eq:GL}
\end{align}
where $\Psi$ is the condensate wave function (order parameter) and ${\bf B} = {\rm rot} \, {\bf A}$ is the magnetic field. The zero-current boundary condition ${\bf n}\cdot \big( - i \mathbf{\nabla} - {\bf A} \big) \Psi =0$  is to be fulfilled at the sample boundary (${\bf n}$ is the unit vector normal to the boundary surface). We stress that in the present work this condition is also used for the in-plane supercurrent, which differs from the case of the periodic boundary conditions in the earlier study~\cite{Cordoba1}. The field satisfies the asymptotic condition $ {\bf B} = {\bf H}$ at infinity, where $ {\bf H}= (0,0,H)$ is the external magnetic field. For the calculations we choose an auxiliary surface at which this asymptotic condition is applied as the boundary condition; this surface has to be sufficiently far from the superconducting sample so that the choice could not influence the results. The flat sample is modelled by a slab with thickness $w$ and two in-plane dimensions $L_a = 50\xi_0$ and $L_b = 75\xi_0$. The condition $L_{a,b} \gg w$ reflects the geometry of a thin film. A sketch of the system is shown in Fig.~\ref{fig1}, where the sample is marked red and the grey colour is for the volume inside the auxiliary surface.

The solution is obtained by the auxiliary time-dependence method, where the GL equations (\ref{eq:GL}) for $\Psi$ and ${\bf B}$ are amended with the first order time derivatives, so that the solution converges to the stationary one at sufficiently large times. The resulting time-dependent GL formalism is solved using the link variable method~\cite{Kato,Cordoba1,Cordoba2}.

Finally, the GL parameter is chosen as $\kappa=0.55$ that corresponds to a type-I material. The temperature is restricted to $T \geq 0.6 T_c$, where the GL theory is expected to be valid at least qualitatively.

\section*{Acknowledgements}

This work was supported by the Brazilian agencies Coordena\c{c}\~ao de Aperfei\c{c}oamento de Pessoal de N\'ivel Superior,  CAPES (Grant No. 223038.003145/2011-00), Conselho Nacional de Ci\^encia e Tecnologia, CNPq (Grants No. 400510/2014-6 and No. 309374/2016-2), and FACEPE (APQ-0936-1.05/15). A.V. acknowledges support from the Russian Science Foundation under the Project 18-12-00429, which was used to study non-local interactions in superconductors. A. S. V. acknowledges support of the joint Russian-Greek Projects RFMEFI61717X0001 and T4$\Delta$P$\Omega$-00031 ``Experimental and theoretical studies of physical properties of low-dimensional quantum nanoelectronic systems''.

\section*{Author contributions statement}

A. Vagov, A. A. Shanenko and W. Y. C\'{o}rdoba-Camacho conceived of the idea of the work; W. Y. C\'{o}rdoba-Camacho and R. M. da Silva developed a numerical code and performed calculations; A. Vagov took the lead in writing the manuscript with support of A. A. Shanenko; J. Albino Aguiar, B. G. Lvov, and A. S. Vasenko contributed to the interpretation of results. All authors provided critical feedback and helped to shape the research and manuscript.

\section*{Competing Interests statement}

The authors declare no competing financial interests.

\newpage

\end{document}